\documentclass[aps,prl,reprint,superscriptaddress]{revtex4-2}
\usepackage{amsmath,amsfonts,amssymb}
\usepackage{graphicx,calc,bm}
\usepackage[breaklinks,colorlinks,urlcolor=blue,citecolor=blue,linkcolor=blue]{hyperref}
\begin{document}
\title{Visualization of skyrmion-superconducting vortex pairs in a chiral magnet-superconductor heterostructure}

\author{Yong-Jie Xie}\thanks{These authors contributed equally to this work}
\affiliation{Beijing National Laboratory for Condensed Matter Physics, Institute of Physics, Chinese Academy of Sciences, Beijing 100190, China}
\affiliation{Center of Materials Science and Optoelectronics Engineering, University of Chinese Academy of Sciences,Beijing 100049, China}
\author{Ang Qian}\thanks{These authors contributed equally to this work}
\affiliation{Beijing National Laboratory for Condensed Matter Physics, Institute of Physics, Chinese Academy of Sciences, Beijing 100190, China}
\affiliation{School of Physical Sciences, University of Chinese Academy of Sciences, Beijing 100049, China}
\author{Bin He}\thanks{These authors contributed equally to this work}
\affiliation{Beijing National Laboratory for Condensed Matter Physics, Institute of Physics, Chinese Academy of Sciences,
Beijing 100190, China}
\affiliation{Center of Materials Science and Optoelectronics Engineering, University of Chinese Academy of Sciences,Beijing 100049, China}
\author{Yu-Biao Wu}\thanks{These authors contributed equally to this work}
\affiliation{Beijing National Laboratory for Condensed Matter Physics, Institute of Physics, Chinese Academy of Sciences, Beijing 100190, China}
\author{Sheng Wang}
\affiliation{Beijing National Laboratory for Condensed Matter Physics, Institute of Physics, Chinese Academy of Sciences, Beijing 100190, China}
\affiliation{School of Physical Sciences, University of Chinese Academy of Sciences, Beijing 100049, China}
\author{Bing Xu}
\affiliation{Beijing National Laboratory for Condensed Matter Physics, Institute of Physics, Chinese Academy of Sciences, Beijing 100190, China}
\author{Guoqiang Yu}
\affiliation{Beijing National Laboratory for Condensed Matter Physics, Institute of Physics, Chinese Academy of Sciences, Beijing 100190, China}
\affiliation{Songshan Lake Materials Laboratory, Dongguan, Guangdong 523808, China}
\author{Xiufeng Han}
\affiliation{Beijing National Laboratory for Condensed Matter Physics, Institute of Physics, Chinese Academy of Sciences, Beijing 100190, China}
\affiliation{Center of Materials Science and Optoelectronics Engineering, University of Chinese Academy of Sciences,Beijing 100049, China}
\affiliation{Songshan Lake Materials Laboratory, Dongguan, Guangdong 523808, China}
\author{X.G. Qiu}\email{xgqiu@iphy.ac.cn}
\affiliation{Beijing National Laboratory for Condensed Matter Physics, Institute of Physics, Chinese Academy of Sciences, Beijing 100190, China}
\affiliation{School of Physical Sciences, University of Chinese Academy of Sciences, Beijing 100049, China}
\affiliation{Songshan Lake Materials Laboratory, Dongguan, Guangdong 523808, China}
\begin{abstract}

Magnetic skyrmions, the topological states possessing chiral magnetic structure with non-trivial topology, have been widely investigated as a promising candidate for spintronic devices. They can also couple with superconducting vortices to form skyrmion-vortex pairs, hosting Majorana zero mode which is a potential candidate for topological quantum computing. A lot of theoretical proposals have been put forward on constructing skyrmion-vortex pairs in heterostructures of chiral magnet and superconductor. Nevertheless, how to generate skyrmion-vortex pairs in a controllable way experimentally remains a significant challenge. We have designed a heterostructure of chiral magnet and superconductor [Ta/Ir/CoFeB/MgO]$_{7}$/Nb in which zero field Néel-type skyrmions can be stabilized and the superconducting vortices can couple with the skyrmions when Nb is in the superconducting state. We have directly observed the formation of skyrmion-superconducting vortex pairs which is dependent on the direction of the applied magnetic field. Our results provide an effective method to manipulate the quantum states of skyrmions with the help of superconducting vortices, which can be used to explore new routines to control the skyrmions for spintronics devices.

\end{abstract}
\maketitle

{\em Introduction.---}
Magnetic skyrmions are small swirling topological defects induced by chiral magnetic interactions or broken inversion symmetry \cite{BogdanovJMMM,BogdanovPRL,RosslerNature,MuhlbauerScience,SkyrmeNuclear,Yu2010Nature}. There are two typical types of magnetic skyrmions: Néel-type \cite{JiangScience,WuAIP,YuNanolett} and Bloch-type \cite{MunzerPRB,NeubauerPRL,OnosePRL,SekiScience} skyrmions, which correspond to different symmetries of chiral Dzyaloshinskii–Moriya interactions \cite{DzyaloshinskyJPCS,MoriyaPR}. Skyrmions can be defined by the topological number (or winding number)  $N=\frac{1}{4\pi}\int {\bf M} \cdot (\frac{\partial {\bf M}}{\partial x} \times \frac{\partial {\bf M}}{\partial y}) dx dy$, which is a measure of the winding of the normalized local magnetization ${\bf M}$ \cite{FertNRM}. Their stabilization and dynamics depend strongly on their topological properties. Owing to their topological property, small size, and high mobility, skyrmions can be manipulated by a very small current density \cite{FertNNT,SampaioNNT}. Skyrmions have been therefore proposed as promising candidates for the next generation, low-power spintronic devices, such as non-volatile information storage and logic devices \cite{BackIOP,NagaosaNNT,ZhangJPCM,KangPROCIEEE}.

\begin{figure*}[t]
	\centering
	\includegraphics[width=0.98\textwidth]{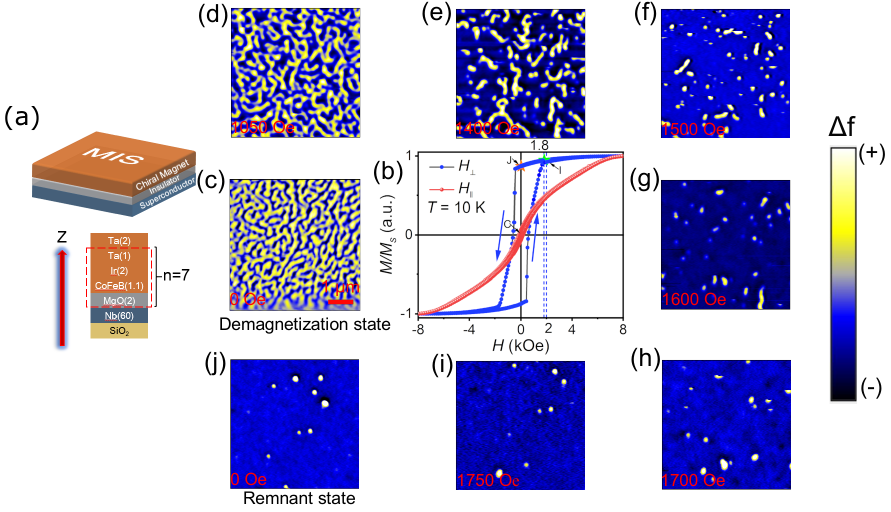}
	\caption{(a) Sample composition: numbers following the layer labels (e.g. Ta(1), Ir(2)) indicate layer thicknesses in nm and there are 7 stacked repeats of the [Ta(1)/Ir(2)/CoFeB(1.1)/MgO(2)] unit. 
	(b) Magnetic hysteresis loops at $T=10$ K for a [Ta(1)/Ir(2)/CoFeB(1.1)/MgO(2)]$_7$ film under perpendicular ($H_{\perp}$) and paralell ($H_{\parallel}$) magnetic fields. The arrows indicate field sweep directions. Individual skyrmions nucleate in the field range of $[1700,2000]$ Oe, marked as blue dash lines.
	(c-j) The MFM images at $T=10$ K in our MIS heterostructure during a perpendicular magnetic field $H_{\perp}=0$ Oe $\rightarrow$ 1750 Oe $\rightarrow$ 0 Oe sweep.  (c) represents the demagnetization state (the initial state before the application of the magnetic field at the origin of (b)) and (j) represents the remnant state (marked by the orange star in (b)). 
	The scale bar is 1 $\mu$m  and color bars indicate the MFM probe resonance shift $\Delta f$ in Hz, proportional to $m_z$. Blue(yellow) color bar indicates $+(-)m_z$. The MFM scanning area is $5\times 5$ $\mu$m$^2$.}
	\label{FIG1}
\end{figure*}

The interaction of skyrmions with another kind of topological excitation, namely superconducting vortices, in heterostructure of chiral magnet and superconductor has been widely discussed recently \cite{HalsPRL,YangPRB}. The broken inversion symmetry of the heterostructure leads to a magnetoelectric coupling that mediates an interaction between skyrmions and superconducting vortices, forming a new type of composite topological excitation referred to as skyrmion-vortex pair (SVP). It is found that non-Abelian Majorana bound states (MBS) exist at the vortex core of SVPs \cite{HalsPRL,YangPRB,PathakPRB,Rex2019PRB}. Therefore, heterostructures of chiral magnet and superconductor have been proposed as a possible platform for hosting Majorana zero modes and exploring topological quantum computation with high fault-tolerance \cite{NothhelferPRB,PershogubaPRB}. It has also been theoretically demonstrated that this kind of heterostructure provides unprecedented tunability of the direction of motion for skyrmions, which facilitates skyrmion guidance in racetrack applications \cite{MenezesPRB}. Experimentally, Kubetzka et al. demonstrated that Fe/Ir on Re substrate could form a skyrmion-superconductor hybrid system \cite{KubetzkaPRM}. Recent experiments on chiral magnet-superconductor heterostructures at applied magnetic fields have confirmed the anti-vortices generated by the stray field of skyrmions based on unique signatures in experimental data as well as numerical simulations \cite{PetrovicPRL}. Nevertheless, in the above mentioned experimental works, the skyrmions are stabilized at relatively high fields so that vortex-vortex interaction is non-negligible and the existence of individual SVP remains to be clarified. Most importantly, how to generate SVPs in a controllable way and directly observe them remains a challenging task. 

We have designed a kind of chiral magnet-superconductor heterostructure (MIS) of 
[Ta/Ir/CoFeB/MgO]$_7$/Nb on Si substrate \cite{sm}. The proximity effect has been suppressed by inserting an insulator layer between the chiral magnet and superconductor, so that skyrmion and vortex interact via stray field only. It is found that in our heterostructures, Néel-type skyrmions can be stabilized at zero field after their nucleation at high fields. Formation of SVPs has been directly observed by magnetic force microscopy (MFM). When a negative magnetic field is applied, the superconducting vortices prefer to locate at the centers of the skyrmions, forming SVPs with enlarged radii. While in a positive field, superconducting vortices are expelled away from the skyrmions. Such isolated SVPs can be an ideal platform to explore Majorana zero mode for topological computation as well as applications based on the manipulation of skyrmions.

\begin{figure}[t]
	\centering
	\includegraphics[width=0.49\textwidth]{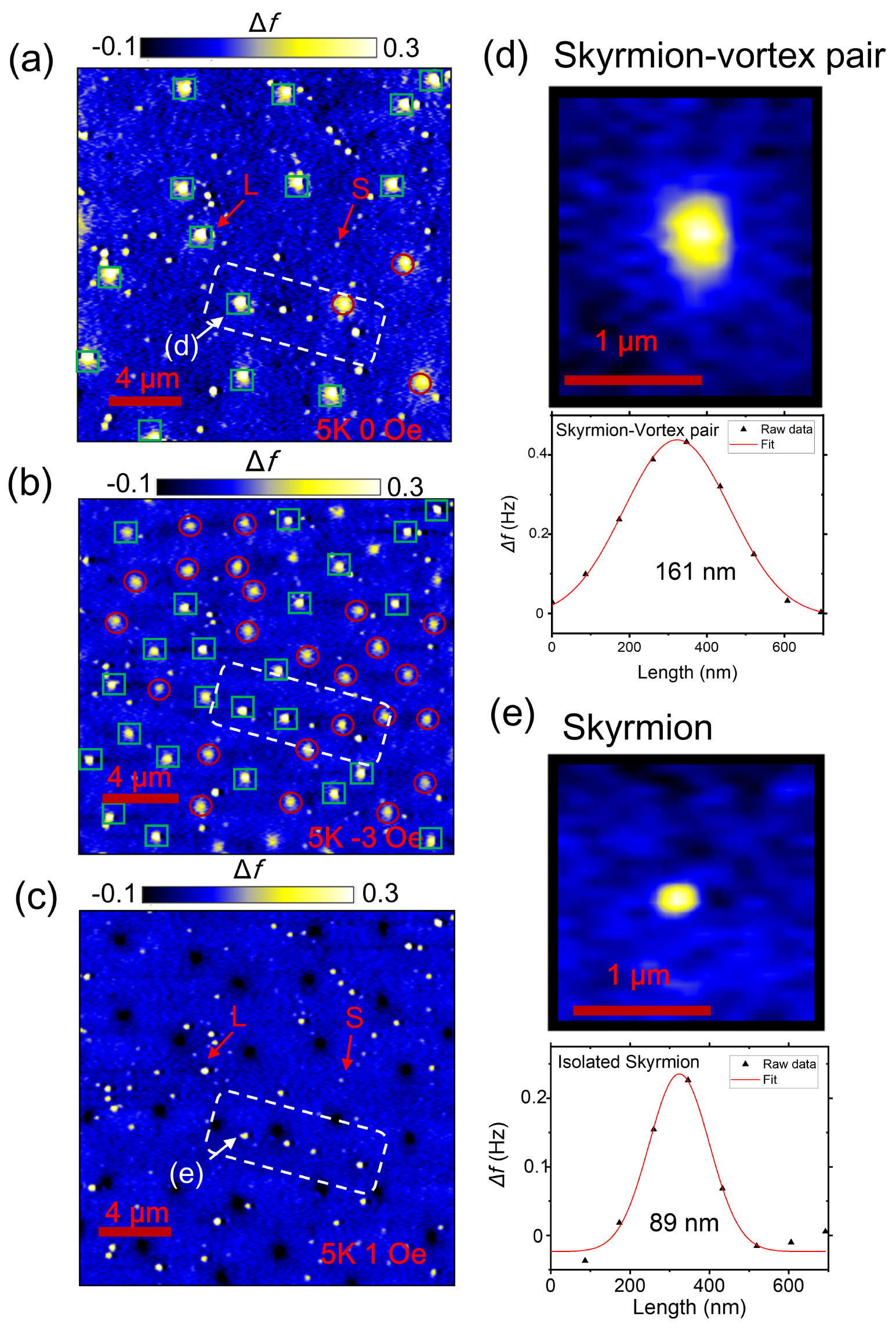}
	\caption{MFM images acquired at $T=5$ K, from $n=$7 MIS heterostructure at different applied magnetic fields of  (a) $H_{\perp}=0$ Oe, (b) $-3$ Oe, (c) 1 Oe. Each image is taken in a field cooling process. The scanning area of (a-c) is $20\times 20$  $\mu$m$^2$. The green squares (red circles) in (a) and (b) represent the vortices induced by the skyrmions with coaxial (non-coaxial) configuration. The SVP and individual skyrmion images with their corresponding MFM profiles in (d) and (e) are enlarged from (a) and (c) respectively, which are marked by white arrows in (a) and (c). Red arrows in (a) and (c) represent large and small skyrmions (marked by `L' and `S') which are used to extract the different profiles in the following numerical calculations.
	 All the scale bars are labeled at left bottom. }
	\label{FIG2}
\end{figure}

{\em Zero-field skyrmions.---}
To obtain skyrmions at zero magnetic field, the magnetic field perpendicular to the sample surface is swept from zero to about 1750 Oe, and then decreased to zero. MFM images have been acquired in several representative fields. From the demagnetization state shown in Fig. \ref{FIG1}(c), the alternating up-and-down (yellow and blue) labyrinthine-like domains with almost half-to-half occupancy have been observed, corresponding to a zero net magnetic moment.

With the increasing external field, the yellow domains become more spatially separated with an enlarged blue region which represents a uniform domain with positive moment $m_z$. When the external field reaches around 1750 Oe (blue dash line region marked in Fig. \ref{FIG1}(b)), the stripe-like domains evolve into stabilized individual skyrmions (yellow round shaped dots), as shown in Fig. \ref{FIG1}(h) and (i). When the field is reduced to zero, stable skyrmions are clearly observed as shown in Fig. \ref{FIG1}(j), 
representing an equilibrium spin configuration.  Because of the inhomogeneity of the sample, the size of the skyrmion is non-uniform (60 $\sim$ 100 nm) (Detailed analysis can be found in Ref. \cite{HeAEM}: Sample series IV). The distance between the isolated skyrmions is of the order of $\mu$m. The large distance between the skyrmions is of a great advantage for the study of the coupling of skyrmions and superconducting vortices.

{\em Direct visualization of SVPs.---}
After the zero-field skyrmions have been formed in the chiral magnet layer at 10 K which is above the superconducting transition temperature $T_c$, a perpendicular magnetic field is applied at the desired strength, and then the sample is cooled down below $T_c$ to 5 K (field cooling (FC)). Fig. \ref{FIG2}(a-c) are the MFM images taken at calibrated magnetic fields of zero, -3 Oe, 1 Oe with FC processes, respectively. A characteristic region is marked by a white dashed rectangle in Fig. \ref{FIG2}(a-c) which represents the same scanning area. 

Compared with the MFM image for the same scanning area at 10 K (see  Fig. S4(b) \cite{sm}), it can be found that besides the isolated skyrmions appear as small yellow dots at fixed positions, there are two other kinds of bright dots with larger radii in Fig. \ref{FIG2}(a) and (b) marked by green squares and red circles, respectively. From Fig. \ref{FIG2}(a-c), we can find the following: (1) When a negative perpendicular magnetic field ($H_{\perp}= -3$ Oe, along $-z$ direction and parallel to the skyrmions core) is applied, both the numbers of green squares and red circles increase, as shown in Fig. \ref{FIG2}(b). The additional green squares are generated from skyrmions in Fig. \ref{FIG2}(a) at the same position. Meanwhile, the additional red circles emerge from the blue background. (2) When a positive external magnetic field $H_{\perp}= 1$ Oe, along $+z$ direction is applied, the green squares in Fig. \ref{FIG2}(a) and (b) turn into skyrmions at the same positions. In the mean time, the red circles are replaced by a triangular array of dark circles as shown in Fig. \ref{FIG2}(c). By comparing with the MFM images of Nb film in Fig. S2, we can attribute the red circles and dark dots to superconducting vortices. The dark circles in Fig. \ref{FIG2}(c) are actually the anti-vortices arising from the reversal of the direction of the magnetic field. It is noticed that the vortices in green square of Fig. \ref{FIG2}(a) always appear at the fixed positions under individual FC processes of different negative fields, in contrast to the random distribution of vortices in bare Nb film as shown in Fig. S4.

We have plotted the profiles of the MFM image of a skyrmion-vortex pair (marked by arrow (d) in Fig. 2(a)) 
and an isolated skyrmion in the same location
(marked by arrow (e) in Fig. 2(c)) (The profiles of the dark round dots in Fig. \ref{FIG2}(c) are given in Fig. S1 \cite{sm}). Both profiles exhibit axial symmetry, with the one of the skyrmions having a radius of approximately 90 nm and that in the green square around 190 nm. It has been reported that the stray field of a skyrmion is strong enough to generate a superconducting vortex in Nb film \cite{PetrovicPRL}. In the mean time, previous theoretical calculations on the interaction between a superconducting vortex and a skyrmion have shown that when minimizing the total free energy is taken into consideration \cite{AndriyakhinaPRB,ApostoloffPRB,BaumardPRB,DahirPRL}, the superconducting vortex will be located at the center of skyrmions, resulting in SVPs with enlarged radii. This is exactly what has been observed here, as marked with green squares in Fig. \ref{FIG2}(a) and (b). 

To further investigate the origin of these MFM signals in green squares, field evolution of the MFM images has been examined by comparing with those on a bare Nb film under different magnetic fields(detailed analysis can be found in \cite{sm}. Firstly, with the increase of the magnetic field from -2 Oe to 2 Oe, no vortex exists at zero field in Nb film when the polarity of vortices changes from negative (yellow dots) to positive (dark dots) as shown in Fig. S4(a). In contrast, skyrmion-induced vortices can be observed at zero field in MIS sample as shown in Fig. S4(b). The reason is that in the case of Nb film, the superconducting vortices are generated by the spatially uniform remnant field which can be compensated by an applied magnetic field. While for the case of MIS sample, some vortices are generated by the stray field of the skyrmions which is not spatially uniform, and cannot be compensated by a uniform applied magnetic field. Secondly, the arrangement of positive and negative vortices keeps the same (triangle lattice) in Nb film, nevertheless in MIS sample, while the vortices form a triangular lattice in a positive magnetic field (dark dots) but form an irregular pattern in a negative field (yellow dots). The detailed experimental data and analysis are provided in the Supplementary Note 3 and Fig. S4 \cite{sm}. It is further noticed that the number of vortices in a negative field in MIS sample (yellow dots) is larger than that in Nb film at the same field, the additional vortices are those marked by green squares in Fig. S4(c). From above, we can attribute the one in green square as SVP, which is a result of one superconducting vortex generated by the stray field of a skyrmion, located at its center. Based on the above analysis, we can conclude that we have directly visualized the formation of SVPs originating from the interaction between skyrmions and superconducting vortices.

\begin{figure}[t]
	\centering
	\includegraphics[width=0.49\textwidth]{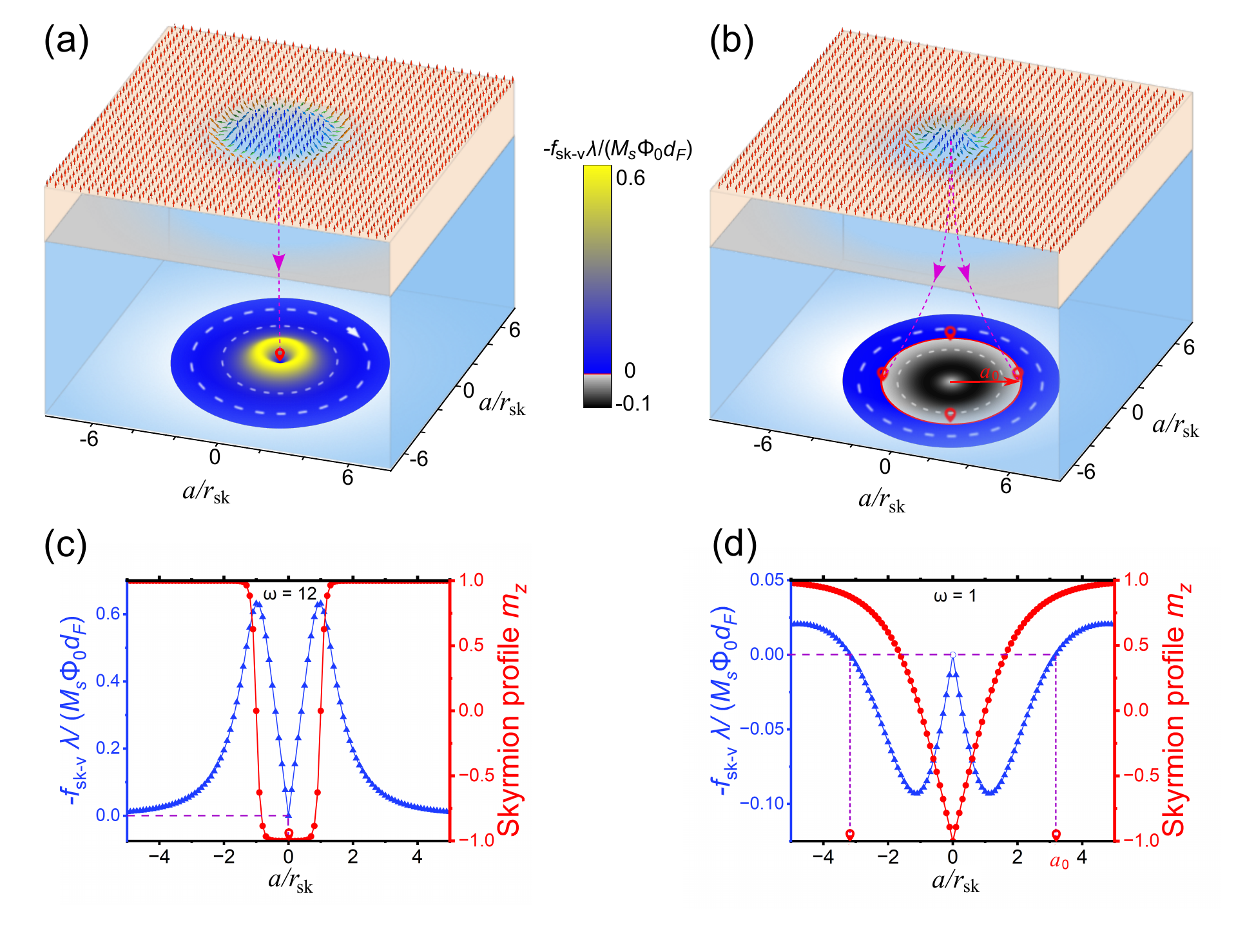}
	\caption{Schematic diagram of the SVP formation in our MIS sample. 
	(a,b) Interaction force between a Néel-type skyrmion with $\omega= 12 (\omega=1)$ in a chiral magnet with a superconducting vortex, SVP with coaxial (non-coaxial) configuration of a skyrmion and a superconducting vortex will be formed.
(c,d) Linecut of the interaction force $-f_{\rm sk-v}$ (blue triangles) profile and the skyrmion profile $m_z$ (red circles) in (a,b).} 
	\label{FIG3}
\end{figure}

{\em Numerical simulation.---}
It is noticed that SVPs tend to be formed on skyrmions with large radii. In order to find the reason, numeric simulations have been carried out using boundary-restrained Maxwell-London equation \cite{AndriyakhinaPRB} to calculate the interaction force between a skyrmion and the superconducting vortex generated by its stray field. Considering the fact that the distance between individual skyrmions is rather large ($\sim$1 $\mu$m), we can neglect the interaction between the skyrmions and treat each SVP as an isolated one. We apply the general form for the field distribution of a superconducting vortex \cite{CarneiroPRB,MenezesPRB} and utilize the same relationship as described in Ref. \cite{AndriyakhinaPRB}. The interaction force between a skyrmion and a superconducting vortex can be expressed as ${\bf f}_{\rm sk-v}={\bf J}_{\rm sk}(a) \times (-\Phi_0){\bf e}_z$, where ${\bf J}_{\rm sk} (a)=\Phi^{-1}_0(\frac{\partial \mathcal{F}_{\rm sk-v}}{\partial a})$ being the derivative of the interaction energy $F_{\rm sk-v}$ with respect to the center-to-center distance $a$, and $\Phi_0$  the flux quantum. The interaction energy and supercurrent distribution strongly depend on the skyrmion polar angle $\theta(a)$ (see Eqs. S1 and S6 in SM \cite{sm}). We use the 360$^\circ$ domain wall ansatz $\theta(a)=\arctan(\frac{\sinh(\omega)}{\sinh(\frac{a\cdot\omega}{r_{\rm sk}})})$ to fit the skyrmion profiles ($r_{\rm sk}$ is the skyrmion radius), and extract different $\omega$, which is defined as the ratio between $r_{\rm sk}$ and the skyrmion wall width $r_{\rm w}$ \cite{WangCP}. The extracted values of $\omega$ for the skyrmions marked by `L' and 'S' in Fig. \ref{FIG2}(a) and (c) are 12 and 1, respectively. The calculated schematic diagram and interaction force distribution for $\omega$=12 and $\omega$=1 are shown in Fig. \ref{FIG3} (the corresponding supercurrent and interaction energy distribution diagram can be found in \cite{sm}. During the calculation, skyrmion chirality $\eta=+1$ has been chosen \cite{ChenCPPS}. It can be seen that when $\omega=12\gg1$ in Fig. \ref{FIG3}(a), the interaction force keeps attractive over a considerable long range, the interaction energy reaches minimum at zero center-to-center distance, i.e., forming a coaxial configuration (a location mark in Fig. \ref{FIG3}(a)). In contrast, when $\omega=1$ in Fig. \ref{FIG3}(b), the interaction force shows repulsive first at zero, and becomes attractive after a specified distance, which indicates that the interaction energy reaches minimum at a limited center-to-center distance $a_0$ (location marks in Fig. \ref{FIG3}(b)). Therefore, SVP with coaxial configuration of a skyrmion and a superconducting vortex will not be formed. These are in good agreement with the observations in Fig. \ref{FIG2}(a) and (c).

{\em Discussions.---}
In a ferromagnet-superconductor heterostructure where inversion symmetry is broken by the interface, the spatially varying stray field of a skyrmion generates a supercurrent in the superconductor to form superconducting vortices. In the calculation by ref. \cite{AndriyakhinaPRB,DahirPRL,MenezesPRB,AndriyakhinaPRB}, skyrmion-vortex interaction via stray fields can lead to the binding of SVPs. It is found that in the ferromagnet-superconductor heterostructure, when the vortices are created directly via the stray field coupling with the ferromagnetic layer, vortices will be pushed away by the skyrmion when the orientation of the field is opposite to that of the skyrmion. This is consistent with our observation that when a positive magnetic field is applied, SVPs will turn into isolated skyrmion as shown in Fig. \ref{FIG2}(c).

For a Néel-type skyrmion, the average magnetization inside the skyrmion radius is opposite to that inside the skyrmion domain wall. Thus, the stray field of skyrmion with larger $\omega$ will be dominated by that inside the radius. On the contrary, for a skyrmion with smaller $\omega$, the stray field inside the radius will be blurred by that from the skyrmion domain wall. Therefore, the strength of the stray field in a skyrmion with larger $\omega$ is stronger than that in a skyrmion with smaller $\omega$. This is consistent with the numeric calculation results that superconducting vortices can only be induced by skyrmions with large $\omega$. Since $\omega$=$r_{\rm sk}/r_{\rm w}$, generally skyrmion with larger $r_{\rm sk}$ tends to have larger $\omega$. As a result, it is easier to induce a coaxial vortex for a skyrmion with a large radius.

{\em Conclusion.---}
We have designed a kind of chiral magnet-superconductor heterostructure of [Ta/Ir/CoFeB/MgO]$_7$/Nb where skyrmions can be stabilized at zero applied magnetic field. The formation of SVPs has been directly observed which is dependent on the direction of the applied magnetic field. The superconducting vortices are located at the centers of the skyrmions under a negative magnetic field, forming SVPs with enlarged radii. While in a positive field, superconducting vortices are expelled away from the skyrmions. The extremely small field operations of manipulating the skyrmion state provide a practical routine for controllable manipulation of skyrmion by means of magnetic field. And the dissipationless superconducting current in this kind of heterostructure can provide a great advantage in spintronics devices based on current-driven skyrmion motion.

{\em Acknowledgments.---}
We thank Prof. X.L. Dong and Dr. Y. Liu for helpful magnetization measurements. This work was supported by the National Key R\&D Program of China (Grant No.2022YFA1403902), the National Natural Science Foundation of China (Grants No. 11974412, 12134017, 12274437 and 12247175), and the Chinese Academy of Sciences (Grant No. XDB33010100).

\bibliographystyle{apsrev4-1}
\bibliography{ref}
\end{document}